\documentclass[11pt]{article}
\usepackage{float,subfigure}
\usepackage{eurosym}
\usepackage{graphicx}
\usepackage{amsmath}
\usepackage{amssymb}
\usepackage{latexsym}
\usepackage{color}
\usepackage{epstopdf}
\usepackage[utf8]{inputenc}
\usepackage[T1]{fontenc}

\providecommand{\openone}{\leavevmode\hbox{\small1\kern-4.3pt\normalsize1}}

\begin{document}

\thispagestyle{empty}

\begin{center}
\vspace{1.8cm}

{\Large \textbf{A comparative study of local quantum Fisher information and local quantum uncertainty in Heisenberg $XY$ model}}

\vspace{1.5cm}

\textbf{A. Slaoui}$^{a}$ {\footnote{%
email: \textsf{abdallahsalaoui1992@gmail.com}}}, \textbf{L. Bakmou}$^{a}$ {\footnote{%
email: \textsf{baqmou@gmail.com}}}, \textbf{M. Daoud}$^{b}$ {\footnote{%
email: \textsf{m\_daoud@hotmail.com}}} and \textbf{R. Ahl Laamara}$^{a,c}$ {%
\footnote{%
email: \textsf{ahllaamara@gmail.com}}}

\vspace{0.5cm}

$^{a}$\textit{LPHE-MS, Faculty of Sciences, Mohammed V University of Rabat,
Rabat, Morocco}\\[0.5em]
$^{b}$\textit{Department of Physics, Faculty of Sciences, University Ibn Tofail, Kénitra, Morocco}\\[0.5em]
$^{c}$\textit{Centre of Physics and Mathematics (CPM), Mohammed V University
of Rabat, Rabat, Morocco}\\[0.5em]

\end{center}
\baselineskip=18pt
\medskip
\vspace{1cm}

\begin{abstract}
Recently, it has been shown that the quantum Fisher information via local observables and via local measurements (i.e., local quantum Fisher information (LQFI)) is a central concept in quantum estimation and quantum metrology and captures the quantumness of correlations in multi-component quantum system [S. Kim et al., Phys. Rev. A. {\bf97}, 032326 (2018)]. This new discord-like measure is very similar to the quantum correlations measure called local quantum uncertainty (LQU). In the present study, we have revealed that LQU is bounded by LQFI in the phase estimation protocol. Also, a comparative study between these two quantum correlations quantifiers is addressed for the quantum Heisenberg $XY$ model. Two distinct situations are considered. The first one concerns the anisotropic $XY$ model and the second situation concerns isotropic $XY$ model submitted to an external magnetic field. Our results confirm that LQFI reveals more quantum correlations than LQU. \\

\end{abstract}

\newpage

\section{Introduction}
Quantum entanglement, as a special type of quantum correlations, is a key ingredient in quantum information science \cite{Einstein1935,Bell1964,Hill1997}. Nowadays, it has been realized that entanglement plays a very important role in many applications in quantum information theory including quantum computing \cite{Nielsen2000}, quantum communications \cite{Braunstein1998,Bouwmeester1997} and quantum key distribution \cite{Ekert1991}. In this sense, their quantification and investigation in multipartite closed and open systems is one of the most fundamental problems in the quantum physics literature. Many different quantification methods have been proposed \cite{Yu2009,Wootters2001} and many studies have investigated the dynamical behavior of entanglement submitted to decoherence effects, which constitutes one of the major challenges for the physical realization of quantum information and computation protocols \cite{Viola1998,Carvalho2004}.\par
On the other hand, recent studies showed that quantum correlations can not only be limited to entanglement, because it represent only one special kind of correlation useful for quantum technology \cite{Datta2008}. It has been shown that some separable quantum states have non-classical correlations that may arise without entanglement and that are also very useful in practical quantum information processing \cite{Lanyon2008,Datta2007}. Additionally, finding physically meaningful and mathematically rigorous quantifiers of the quantum correlation beyond entanglement have attracted an impressive amount of attention and efforts. Historically, entropic quantum discord (QD), first introduced by Ollivier and
Zurek \cite{Ollivier2001} and later proposed by Henderson and Vedral \cite{Henderson2001}, is the widely used quantum correlations quantifier and several works were dedicated to this class of measure. Streltsov and Zurek \cite{Streltsov2013} showed that the measurement results cannot be communicated perfectly by classical means if a measurement device is in a nonclassical state. This means that the lost of information occurs even when the measurement apparatus is not entangled with the system and the amount of this lost information turns out to be exactly the quantum discord. Moreover, Werlang et al.\cite{Werlang2009}, showed that QD is more robust than entanglement under the same Markovian environment conditions for dissipative systems. Quantum discord is defined as the difference between two classically-equivalent expressions of the mutual information: the quantum mutual information and the local measurement-induced quantum mutual information. The analytical expressions of QD have been found only for some special types of states and the situation becomes more complicated for general quantum states. This is due to an optimization procedure for the conditional entropy over all local generalized measurements which is difficult to perform. The computation of QD is NP-complete \cite{Huang2014}. These difficulties led Dakic et al., to propose an alternative formulation of QD which is called “the geometric measure of quantum discord (GMQD)”. It is the minimum Hilbert-Schmidt distance between the given state and the zero discord state \cite{Dakic2010}. From an analytical point of view, the calculation of this geometric measure requires a simpler minimization process compared to the entropic QD \cite{Daoud2012,Daoud2015}. Despite this remarkable feature, the GMQD cannot be considered as a faithful quantifier of non-classical correlations \cite{Piani2012}.\par
To overcome such difficulties and problems, now tools were proposed to quantify the non-classical correlations. In this sense, Girolami et al., \cite{Girolami2013} have introduced the concept of local quantum uncertainty (LQU) as a discord-like measure of quantum correlation. It is defined as the minimum uncertainty induced by measurement of a single local observable on a quantum state using the notion of Wigner-Yanase skew information \cite{Wigner1963}. This measure satisfies the full physical requirements of a measure of quantum correlations. Beyond its importance as a quantum correlation quantifier, LQU is also deeply related to quantum Fisher information (QFI) in the context of quantum metrology. Indeed, it has been proven that in the case of  unitary dynamics of a state $\rho$, i.e, ${\rho _\theta } = U_\theta ^\dag \rho {U_\theta }$ with ${U_\theta } = {e^{i{H}\theta }}$, the skew information is majorized by the QFI associated with the phase parameter and the quantum analog of the classical Cramer-Rao inequality can be written in term of the skew information \cite{Luo12003,Luo22003}. Therefore, the LQU of a bipartite mixed probe state guarantees a minimum precision quantified by the QFI in the optimal phase estimation protocol \cite{Girolami2013}. Recently, the concept of local quantum Fisher information (LQFI) was introduced in \cite{Kim2018} to characterize quantum correlations in terms of QFI. This reliable quantum discord-like quantifier, is based on minimising QFI over a local maximally informative observable associated with one of the subsystems. Furthermore, LQFI offers a promising tool to understand the role of quantum correlations other than entanglement in enhancing the precisions and the efficiency of quantum metrology protocols.\par
Since LQU and LQFI are both based on the notion of quantum uncertainty, it is of interest to study the relationship and interplay between them. This is the main issue that we develop in this work. We give the relation between LQU and LQFI for qubit-qudit systems. This paper is structured as follows. In section 2, we provide an overview on LQFI and LQU  to quantify the degree of quantum correlations contained in a bipartite quantum system. We show that LQU is majorized by LQFI in any metrological task of the phase estimation. In section 3, we analyze the behavior of the LQU and compare it to the behavior of the LQFI in Heisenberg $XY$ model. We consider two special situations \cite{Zhang2007}. The first one concerns the anisotropic $XY$ model and the second situation concerns the isotropic $XY$ model submitted to an external magnetic field. Concluding remarks close this paper.
\section{Quantifiers of non-classical correlations by quantum uncertainty}

\subsection{Local quantum Fisher information}
In quantum estimation theory, the quantum Fisher information (QFI) is recognized as the most widely used quantity for characterizing the ultimate accuracy in parameter estimation scenarios \cite{Helstrom1976,Kay1993,Genoni2011}. More recently, many efforts have been made in evaluating the dynamics of QFI to establish the relevance of quantum entanglement in quantum metrology \cite{Blondeau2017,Giovannetti2004}. It has been demonstrated that, in the unitary processes, entanglement leads to a notable improvement of the accuracy of parameter estimation \cite{Huelga1997,Blondeau2016}. In this context, it is natural to wonder whether quantum correlations beyond entanglement can be related to precision in quantum metrology protocols. Usually, for an arbitrary quantum state $\rho _\theta$ that depends on the parameter $\theta$, we can define the QFI as \cite{Luo12003,Ye2018,Paris2009}
\begin{equation}
\mathcal{F}\left( \rho _\theta \right)= \frac{1}{4}{\rm Tr}\left[ \rho _\theta {L_\theta}^2 \right],
\end{equation}
where the symmetric logarithmic derivative $L_\theta$  is defined as the solution of the equation
\begin{equation}
\frac{{\partial {\rho _\theta }}}{{\partial \theta }} = \frac{1}{2}\left( {{L_\theta }{\rho _\theta } + {\rho _\theta }{L_\theta }} \right).
\end{equation}
The parametric states $\rho_\theta$ can be obtained from an initial probe state $\rho $ subjected to a unitary transformation ${U_\theta } = {e^{i{ H}\theta }}$ dependent on $\theta$ and generated by a Hermitian operator $ H$, i.e., ${\rho _\theta } = U_\theta ^\dag \rho {U_\theta }$. In this case ${\cal F}\left( {{\rho _\theta }} \right)$, that we denote by ${\cal F}\left( {\rho , H} \right)$, is given by
\begin{equation}
{\cal F}\left( {\rho , H} \right) = \frac{1}{2}\sum\limits_{i \ne j} {\frac{{{{\left( {{p_i} - {p_j}} \right)}^2}}}{{{p_i} + {p_j}}}{{\left| {\langle {\psi _i}|{ H}\left| {{\psi _j}} \right\rangle } \right|}^2}},
\end{equation}
where we used the spectral decomposition of $\rho$, i.e., $\rho  = \sum\limits_{i = 1} {{p_i}\left| {{\psi _i}} \right\rangle \langle {\psi _i}|} $ with ${p_i} \ge 0$ and $\sum\limits_{i = 1} {{p_i}}  = 1$. The QFI is $\theta$-independent. Now we consider an $2 \times d$ bipartite quantum state ${\rho _{AB}}$ in the Hilbert space $\cal H = H_{\rm A} \otimes H_{\rm B}$. We assume that the dynamics of the first part is governed by the local phase shift transformation ${e^{ - i\theta {H_A}}}$, with ${H_A} \equiv {H_a} \otimes {\mathbb{I}_B}$ is the local Hamiltonian. In this case, QFI reduces to local quantum Fisher information (LQFI). It writes \cite{Bera2014}
\begin{equation}
\mathcal{F}\left( {\rho ,{H_A}} \right) = {\rm Tr}\left( {\rho {H_A}^2} \right) - \sum\limits_{i \ne j} {\frac{{2{p_i}{p_j}}}{{{p_i} + {p_j}}}{{\left| {\left\langle {{\psi _i}} \right|{H_A}\left| {{\psi _j}} \right\rangle } \right|}^2}}. \label{QFI}
\end{equation}
Local quantum Fisher  information was introduced to deal with pairwise quantum correlations of discord type \cite{Kim2018}. This quantifier enables us to gain a deeper insight on how quantum correlations are instrumental in setting metrological precision. It has the desirable properties that any good quantum correlation quantifiers should satisfy. Indeed, it is non-negative and vanishes for zero discord bipartite states (classically correlated states). It is invariant under any local unitary operation and coincides with the geometric discord for pure quantum states \cite{Kim2018}. The quantification of quantum correlations in terms of local quantum Fisher information $\mathcal{Q}\left( \rho  \right)$ is defined as the minimum quantum Fisher information over all local Hamiltonians ${H_A}$ acting on the $A$-part \cite{Kim2018}
\begin{equation}
\mathcal{Q}\left( \rho  \right) = \mathop {{\rm min}}\limits_{{H_A}} \mathcal{F}\left( {\rho ,{H_A}} \right).
\end{equation}
The general form of a local Hamiltonian is ${H_a} =  {\vec\sigma}  .{\vec r} $ with $\left| {\vec r} \right| = 1$ and ${\vec \sigma}  = \left(  {{\sigma _x}\equiv{\sigma _1},{\sigma _y}\equiv{\sigma _2},{\sigma _z}\equiv{\sigma _3}} \right)$ are the usual Pauli matrices. It can be seen that ${\rm Tr}\left( {\rho {H_A}^2} \right)=1$ and the second term in the equation (\ref{QFI}) can be expressed as
\begin{align}
\sum\limits_{i \ne j} {\frac{{2{p_i}{p_j}}}{{{p_i} + {p_j}}}{{\left| {\langle {\psi _i}|{H_A}\left| {{\psi _j}} \right\rangle } \right|}^2}} & = \sum\limits_{i \ne j} {\sum\limits_{l,k=1}^3 {\frac{{2{p_i}{p_j}}}{{{p_i} + {p_j}}}} } \langle {\psi _i}|{\sigma _l} \otimes \mathbb{I}_B\left| {{\psi _j}} \right\rangle \langle {\psi _j}|{\sigma _k} \otimes \mathbb{I}_B\left| {{\psi _i}} \right\rangle  \\\notag&= {{\vec r}^{\dag} }.{\rm M}.{\vec r}
\end{align}
where the elements of the $3\times3$ symmetric matrix ${\rm M}$ are given by
\begin{equation}
{{\rm M}_{lk}} = \sum\limits_{i \ne j} {\frac{{2{p_i}{p_j}}}{{{p_i} + {p_j}}}\langle {\psi _i}|{\sigma _l} \otimes \mathbb{I}_B\left| {{\psi _j}} \right\rangle \langle {\psi _j}|{\sigma _k} \otimes \mathbb{I}_B\left| {{\psi _i}} \right\rangle }. \label{Mlk}
\end{equation}
To minimize $ \mathcal{F}\left( {\rho ,{H_A}} \right)$, it is necessary to maximize the quantity ${{\vec r}^\dag }.{\rm M}.{\vec r}$ over all unit vectors ${\vec r}$. The maximum value coincides with the maximum eigenvalue of ${\rm M}$. Hence, the minimal value of local quantum Fisher information $\mathcal{Q}\left( \rho  \right)$ is
\begin{equation}
\mathcal{Q}\left( \rho  \right) = 1 - \lambda _{\rm max}\left({\rm M}\right), \label{LQFI}
\end{equation}
where $\lambda _{\rm max}$ denotes the maximal eigenvalue of the symmetric matrix ${\rm M}$ defined by (\ref{Mlk}).
\subsection{Local quantum uncertainty}
Unlike the classic case, due to the probabilistic character of quantum mechanics, two noncommuting observables cannot be simultaneously measured with arbitrary precision and the uncertainty relation imposes a fundamental limit on the precision. Usually, the total uncertainty due to a single observable $H$ measurement in a quantum state $\rho$ is quantified by the variance as ${\rm Var}\left( {\rho ,H} \right) = {\rm Tr}\left( {\rho {H^2}} \right) - {\left( {\rm Tr}\left( {\rho H} \right)\right)^2}$. This uncertainty may exhibit contributions of quantum and classical nature when the statistical error in its measurement is due to classical ignorance about the classical mixing in $\rho$. To quantify only the quantum part of the variance, Wigner and Yanase introduced the notion of skew information as a good quantifier of this uncertainty. It is defined as \cite{Wigner1963}
\begin{equation}
\mathcal{I}(\rho, H)=-\frac{1}{2}{\rm
    Tr}([\sqrt{\rho}, H]^{2}).
\end{equation}
For a bipartite quantum state $\rho\equiv{\rho_{AB}}$, a discord-like measure of quantum correlations was recently introduced by Girolami et al \cite{Girolami2013}. This is called local quantum uncertainty and it is defined as the minimum skew information attainable with a single local measurement. It is given by \cite{Karpat2014,Guo2015}
\begin{equation}
\mathcal{U}(\rho) \equiv \min_{H_A}\mathcal{I}(\rho,H_A \otimes
\mathbb{I}_B), \label{LQU}
\end{equation}
where $H_A$ is a Hermitian operator acting on the subsystem $A$ admitting a nondegenerate spectrum and $\mathbb{I}_B$ is the identity operator acting on the subsystem $B$. For pure bipartite states, the local quantum uncertainty reduces to the linear entropy of entanglement and vanishes for classically correlated states \cite{Girolami2013}. This measure is directly linked to the concept of quantum discord and leads to an entire class of bona fide measures of non-classical correlations. It is also effortlessly computable contrarily to some other measures for which closed analytical expressions are not always easy obtainable. The closed form of the local quantum uncertainty for $2\times d$ quantum systems is
\begin{equation}
\mathcal{U}(\rho) = 1 - {\rm max}\{ \xi_1, \xi_2, \xi_3\}, \label{Lqu}
\end{equation}
where $\xi_{1}, \xi_2$ and $\xi_3$ are the eigenvalues of the $3\times3$ matrix $W$ whose matrix elements are defined by
\begin{equation}\label{w-elements}
\omega_{ij} \equiv  {\rm
    Tr}\{\sqrt{\rho}(\sigma_{i}\otimes
\mathbb{I}_B)\sqrt{\rho}(\sigma_{j}\otimes \mathbb{I}_B)\},
\end{equation}
with $i,j = 1, 2, 3$ and the $\sigma_{i}$ represent the Pauli matrices. For the so-called X states, containing the non-zero entries only along the diagonal and anti-diagonal (states with a visual appearance like the letter $X$ \cite{Ali2010,Vinjanampathy2011}), the analytical expression of local quantum uncertainty was derived in Refs \cite{Slaoui2018,Slaoui22018}. Using the identity resolution of the orthonormal basis $\left\{ {\left| {{\psi _i}} \right\rangle } \right\}$, i.e., $\sum\limits_i {\left| {{\psi _i}} \right\rangle \left\langle {{\psi _i}} \right| = \mathbb{I}}$, it is simple to verify that \cite{Luo12003}
\begin{equation}
\mathcal{I}(\rho, H) = \frac{1}{2}{\sum\limits_{i,j} {{{\left( {\sqrt{p_i} - \sqrt{p_j}} \right)}^2}\left| {\left\langle {{\psi _i}} \right|H\left| {{\psi _j}} \right\rangle } \right|} ^2},
\end{equation}
and
\begin{equation}
{\cal F}\left( {\rho , H} \right) = \frac{1}{2}{\sum\limits_{i,j} {\left( {1 + \frac{{2 \sqrt{p_ip_j}}}{{{p_i}+{p_j}}}} \right){{\left( {\sqrt{p_i} - \sqrt{p_j}} \right)}^2}\left| {\left\langle {{\psi _i}} \right|H\left| {{\psi _j}} \right\rangle } \right|} ^2}.
\end{equation}
The skew information and QFI satisfy the following inequality
\begin{equation}
\mathcal{I}(\rho, H) \le {\cal F}\left( {\rho , H} \right) \le 2 \mathcal{I}(\rho, H),
\end{equation}
from which one gets
\begin{equation}
\mathcal{U}\left(\rho\right) \le \mathcal{Q}\left( \rho  \right) \le 2\mathcal{U}(\rho). \label{UQ2U}
\end{equation}
In quantum metrology, the parameter $\theta$ can be estimated through an (unbiased) estimator $\hat\theta$ and the limit of the precision of their measurement is usually framed by the quantum Cramér-Rao bound which writes \cite{Paris2009}
\begin{equation}
{\rm Var}\left( {\hat \theta } \right) \ge \frac{1}{{n {\cal F}\left( {\rho , H} \right)}},
\end{equation}
where $n$ is the number of times the estimation protocol is repeated. According to Cramér–Rao theorem, more precision is obtained for small variance. It is clear that the inverse of the QFI depicts the lower error limit in statistical estimation of an unknown parameter. It should also be noted that, in the unitary evolution, LQU is majorized by QFI, i.e.,
\begin{equation}
\mathcal{U}(\rho) \le \mathcal{I}\left( {\rho ,H} \right) = \mathcal{I}\left( {{\rho _\theta },H} \right) \le {\cal F}\left( {\rho ,H} \right).
\end{equation}
For $n=1$, the parameter precision can be bound by LQU and by LQFI as
\begin{equation}
{\rm Var}\left( {\hat \theta } \right)_{\min} \le \frac{1}{{\mathcal{U}\left(\rho\right)}}, \hspace{1cm}{\rm and}\hspace{1cm}{\rm Var}\left( {\hat \theta } \right)_{\min} \le \frac{1}{{\mathcal{Q}\left(\rho\right)}}.
\end{equation}

\section{Non-classical correlations in the quantum Heisenberg XY model}
The Heisenberg Hamiltonian for a chain of $N$ qubits can be written as \cite{Korepin1993,Wang2002,Kamta2002}
\begin{equation}
H = \frac{1}{4}\sum\limits_{i = 1}^N {\left( {{J_x}\sigma _i^x\sigma _{i + 1}^x + {J_y}\sigma _i^y\sigma _{i + 1}^y + {J_z}\sigma _i^z\sigma _{i + 1}^z} \right)}, \label{H}
\end{equation}
where $\sigma _i^\alpha \left( {\alpha  = x,y,z} \right)$ are Pauli matrices satisfying the usual commutation relations and ${J_\alpha }$ are the coupling constants of the model. For ${J_x} \ne {J_y} \ne {J_z}$, the model is called the anisotropic Heisenberg $XYZ$ model. In the case where ${J_x} = {J_y} \ne {J_z}$, the Hamiltonian describes the partial anisotropic Heisenberg $XXZ$ model. Finally, for  ${J_x} = {J_y} = {J_z}$ this corresponds to the isotropic Heisenberg $XXX$ model. The interaction in the chain is said to be antiferromagnetic for ${J_\alpha }$ positive and ferromagnetic for ${J_\alpha }$ negative. We assume periodic boundary conditions, so that the $(N+1)$th site is identified with the first site. Recently, several studies have been conducted on Heisenberg spin systems to investigate their entanglement properties at certain temperatures. In this context, an interesting type of entanglement called thermal entanglement was introduced and analyzed in Heisenberg models $XYZ$ \cite{Guo2017}, $XXZ$ \cite{Li2008} and $XXX$ \cite{Yong2007}. In this work, we employ the concepts of LQFI and LQU to investigate the non-classical correlations in the $XY$ (${J_z} = 0$) anisotropic model and the $XY$ (${J_z} = 0$)  isotropic model in external magnetic field. For a system in thermal equilibrium, the corresponding state is described by the density operator $\rho  = {{\exp \left( { - \beta H} \right)} \mathord{\left/
        {\vphantom {{\exp \left( { - \beta H} \right)} Z}} \right.
        \kern-\nulldelimiterspace} Z}$ where $Z = {\rm Tr}\left[ {\exp \left( - \beta H\right)} \right]$ is the partition function and $\beta={1 \mathord{\left/
{\vphantom {1 {kT}}} \right.
\kern-\nulldelimiterspace} {kT}}$ ($k$ is the Boltzmann constant and $T$ is the temperature) and $H$ is the system Hamiltonian. In the numerical calculations, we set $k=1$.

\subsection{ Two-qubit anisotropic  $XY$ model}
The two-qubit anisotropic $XY$ model is obtained from (\ref{H}) for $N = 2$ by setting zero the coupling constant on the $z$-axis ($J_z = 0$). In this case, using the raising and lowering operators $\sigma _i^ \pm  = \frac{1}{2}\left( {\sigma _i^x \pm i\sigma _i^y} \right)$, the Hamiltonian (\ref{H}) becomes
\begin{equation}
H = J\left( {\sigma _1^ + \sigma _2^ -  + \sigma _2^ - \sigma _1^ + } \right) + J \gamma \left( {\sigma _1^ + \sigma _2^ +  + \sigma _1^ - \sigma _2^ - } \right),
\end{equation}
with $J = \frac{{{J_x} + {J_y}}}{2}$ and $\gamma  = \frac{{{J_x} - {J_y}}}{{J_x} + {J_y}}$. Without loss of generality, we take $J=1$.  The parameter $\gamma$ is the anisotropic parameter; it is zero (${J_x} = {J_y}$) for the isotropic $XX$ model and $\pm 1$ for the Ising model. The Hamiltonian $H$ satisfies the eigenvalues equations:
\begin{equation}
H\left| {{\psi ^ \pm }} \right\rangle  =  \pm \left| {{\psi ^ \pm }} \right\rangle, \hspace*{1cm} H\left| {{\chi ^ \pm }} \right\rangle  =  \pm  \gamma\left| {{\chi ^ \pm }} \right\rangle ,
\end{equation}
where $\left| {{\psi ^ \pm }} \right\rangle  = \frac{1}{{\sqrt 2 }}\left( {\left| {01} \right\rangle  \pm \left| {10} \right\rangle } \right)$ and $\left| {{\chi ^ \pm }} \right\rangle  = \frac{1}{{\sqrt 2 }}\left( {\left| {00} \right\rangle  \pm \left| {11} \right\rangle } \right)$ are the states of Bell maximally entangled. In the standard basis $\left\{ {\left| {00} \right\rangle ,\left| {01} \right\rangle ,\left| {10} \right\rangle ,\left| {11} \right\rangle } \right\}$, the density matrix of this system is given by
\begin{equation}
\rho  = \frac{1}{Z}\left( {\begin{array}{*{20}{c}}
    {\cosh \left( {\gamma \beta } \right)}&0&0&{ - \sinh \left( {\gamma \beta } \right)}\\
    0&{\cosh \left( {\beta } \right)}&{ - \sinh \left( {\beta } \right)}&0\\
    0&{ - \sinh \left( {\beta } \right)}&{\cosh \left( {\beta } \right)}&0\\
    { - \sinh \left( {\gamma \beta } \right)}&0&0&{\cosh \left( {\gamma \beta } \right)}
    \end{array}} \right), \label{gh1}
\end{equation}
where $Z = {\rm Tr}\left[ {\exp \left( { - H\beta } \right)} \right] = 2\left( {\cosh \left( {\beta } \right) + \cosh \left( {\gamma \beta } \right)} \right)$ is the partition function. The eigenvalues of the density matrix $\rho$ (\ref{gh1}) are given by
\begin{equation}
{p_1} = \frac{{{e^{ - \beta }}}}{Z},\hspace{1.5cm}{p_2} = \frac{{{e^\beta }}}{Z},\hspace{1.5cm}{p_3} = \frac{{{e^{ - \gamma \beta }}}}{Z},\hspace{1.5cm}{p_4} = \frac{{{e^{\gamma \beta }}}}{Z},
\end{equation}
and the corresponding eigenstates are respectively given by
\begin{equation}
\begin{array}{*{20}{l}}
{\left| {{\psi _1}} \right\rangle  = \frac{1}{{\sqrt 2 }}\left( {\left| {01} \right\rangle  + \left| {10} \right\rangle } \right),\hspace{1cm}\left| {{\psi _2}} \right\rangle  = \frac{1}{{\sqrt 2 }}\left( {\left| {10} \right\rangle  - \left| {01} \right\rangle } \right),}\\
{\left| {{\psi _3}} \right\rangle  = \frac{1}{{\sqrt 2 }}\left( {\left| {00} \right\rangle  + \left| {11} \right\rangle } \right),\hspace{1cm}\left| {{\psi _4}} \right\rangle  = \frac{1}{{\sqrt 2 }}\left( {\left| {11} \right\rangle  - \left| {00} \right\rangle } \right).}
\end{array}
\end{equation}
To obtain the explicit expression of LQFI, one computes first the elements of the matrix $M$. From (\ref{Mlk}), it is easy to check that their off-diagonal elements are equal to zero and the diagonal elements are given as
\begin{equation}
{{\rm{M}}_{11}} = {\rm{sech}^2}{\left( {\frac{{\left( {1 - \gamma } \right)\beta }}{2}} \right)},\hspace{1cm} {{\rm{M}}_{22}} = {\rm{sech}^2}{\left( {\frac{{\left( {1 + \gamma } \right)\beta }}{2}} \right)}, \hspace{1cm} {{\rm M}_{33}} = {\rm{sech}}\left( {\beta } \right){\rm{sech}}\left( {\gamma \beta } \right).
\end{equation}
It is simple to verify that
\begin{equation}
{{\rm M}_{11}} - {{\rm M}_{33}} = \frac{{\cosh \left( {\left( {1 + \gamma } \right)\beta } \right) - 1}}{{\cosh \left( \beta  \right)\cosh \left( {\gamma \beta } \right)\left[ {\cosh \left( {\left( {1 - \gamma } \right)\beta } \right) + 1} \right]}},\label{M11-M33}
\end{equation}
and
\begin{equation}
{{\rm M}_{22}} - {{\rm M}_{33}} = \frac{{\cosh \left( {\left( {1 - \gamma } \right)\beta } \right) - 1}}{{\cosh \left( \beta  \right)\cosh \left( {\gamma \beta } \right)\left[ {\cosh \left( {\left( {1 + \gamma } \right)\beta } \right) + 1} \right]}}.\label{M22-M33}
\end{equation}
From equations (\ref{M11-M33}) and (\ref{M22-M33}), it is clear that ${{\rm M}_{33}} \le {{\rm M}_{11}}$ and ${{\rm M}_{33}} \le {{\rm M}_{22}}$. Therefore, the analytical expression of LQFI is
\begin{equation}
\mathcal{Q}\left( \rho  \right)= \left\{ \begin{array}{l}
1 - {{\rm M}_{22}}\hspace{1cm} {\rm for}\hspace{0.6cm} - 1 \le \gamma\le 0\\
1 - {{\rm M}_{11}} \hspace{1cm}{\rm for} \hspace{1cm}0 \le \gamma  \le 1
\end{array} \right.
\end{equation}
On the other hand, to determine the analytic expression of LQU given by (\ref{Lqu}), it is necessary to compute the matrix elements given by (\ref{w-elements}). After straightforward calculation, the LQU is
\begin{equation}
\mathcal{U}(\rho) = 1 - {\rm max}\{ \omega_{11}, \omega_{22}, \omega_{33}\},
\end{equation}
where
\begin{equation}
{\omega_{11}} = \frac{{\left( {\cosh \left( \beta  \right) + 1} \right)\left( {\cosh \left( {\beta \gamma } \right) + 1} \right) + \sinh \left( {\beta \gamma } \right)\sinh \left( \beta  \right)}}{{\left( {\cosh \left( {\beta \gamma } \right) + \cosh \left( \beta  \right)} \right)\sqrt {\left( {\cosh \left( \beta  \right) + 1} \right)\left( {\cosh \left( {\beta \gamma } \right) + 1} \right)} }}, \label{w111}
\end{equation}
\begin{equation}
{\omega_{22}} = \frac{{\left( {\cosh \left( \beta  \right) + 1} \right)\left( {\cosh \left( {\beta \gamma } \right) + 1} \right) - \sinh \left( {\beta \gamma } \right)\sinh \left( \beta  \right)}}{{\left( {\cosh \left( {\beta \gamma } \right) + \cosh \left( \beta  \right)} \right)\sqrt {\left( {\cosh \left( \beta  \right) + 1} \right)\left( {\cosh \left( {\beta \gamma } \right) + 1} \right)} }},\label{w122}
\end{equation}
and
\begin{equation}
{\omega_{33}} = \frac{2}{{ {\cosh \left( {\beta \gamma } \right) + \cosh \left( \beta  \right)}}}.\label{w133}
\end{equation}
To compare the eigenvalues $\omega_{11}$, $\omega_{22}$ and $\omega_{33}$, we analyse the sign of the following quantities:
\begin{equation}
{\mathop{\rm sign}\nolimits} \left( {{\omega_{11}} - {\omega_{33}}} \right) = {\mathop{\rm sign}\nolimits} \left( {\cosh \left( {\left( {1 + \gamma } \right)\beta } \right) - 1 + {{\left( {\sqrt {\cosh \left( \beta  \right) + 1}  - \sqrt {\cosh \left( {\gamma \beta } \right) + 1} } \right)}^2}} \right),
\end{equation}
and
\begin{equation}
{\mathop{\rm sign}\nolimits} \left( {{\omega_{22}} - {\omega_{33}}} \right) = {\mathop{\rm sign}\nolimits} \left( {\cosh \left( {\left( {1 - \gamma } \right)\beta } \right) - 1 + {{\left( {\sqrt {\cosh \left( \beta  \right) + 1}  - \sqrt {\cosh \left( {\gamma \beta } \right) + 1} } \right)}^2}} \right).
\end{equation}
We note that ${\omega_{33}} \le {\omega_{11}}$ and ${\omega_{33}} \le {\omega_{22}}$. Thus, to determine the expression of LQU, one should find the conditions under which ${\omega_{11}} \le {\omega_{22}}$ or ${\omega_{22}} \le {\omega_{11}}$. Finally, we obtain
\begin{equation}
\mathcal{U}\left( \rho  \right)  = \left\{ \begin{array}{l}
1 - {\omega_{11}}\hspace{1cm}{\rm when}\hspace{1cm}0 \le \gamma  \le 1\\
1 - {\omega_{22}}\hspace{1cm}{\rm when}\hspace{0.6cm} - 1 \le \gamma  \le 0
\end{array} \right.
\end{equation}

\begin{figure}[H]
\centering
\begin{subfigure}[]
    \centering
    \includegraphics[width=3in]{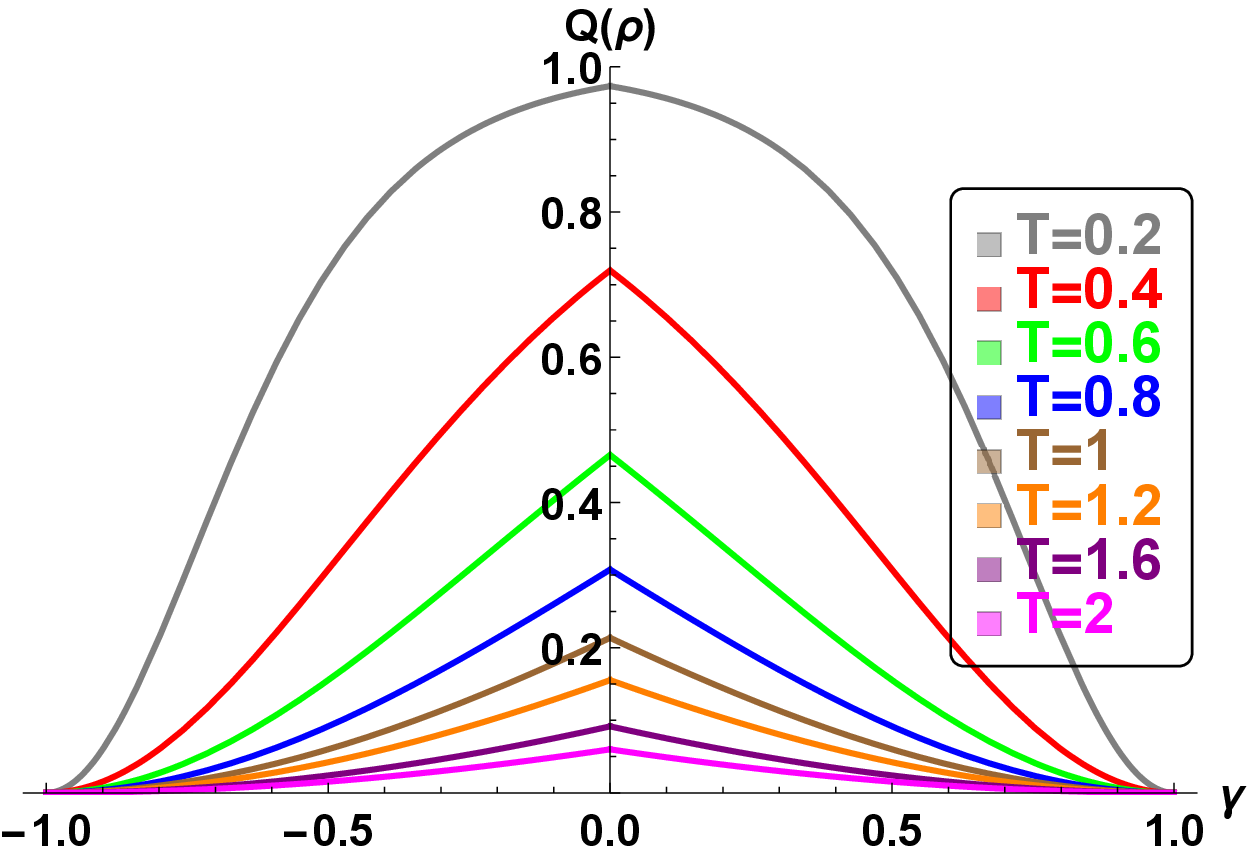}
\end{subfigure}
~
\begin{subfigure}[]
    \centering
    \includegraphics[width=3in]{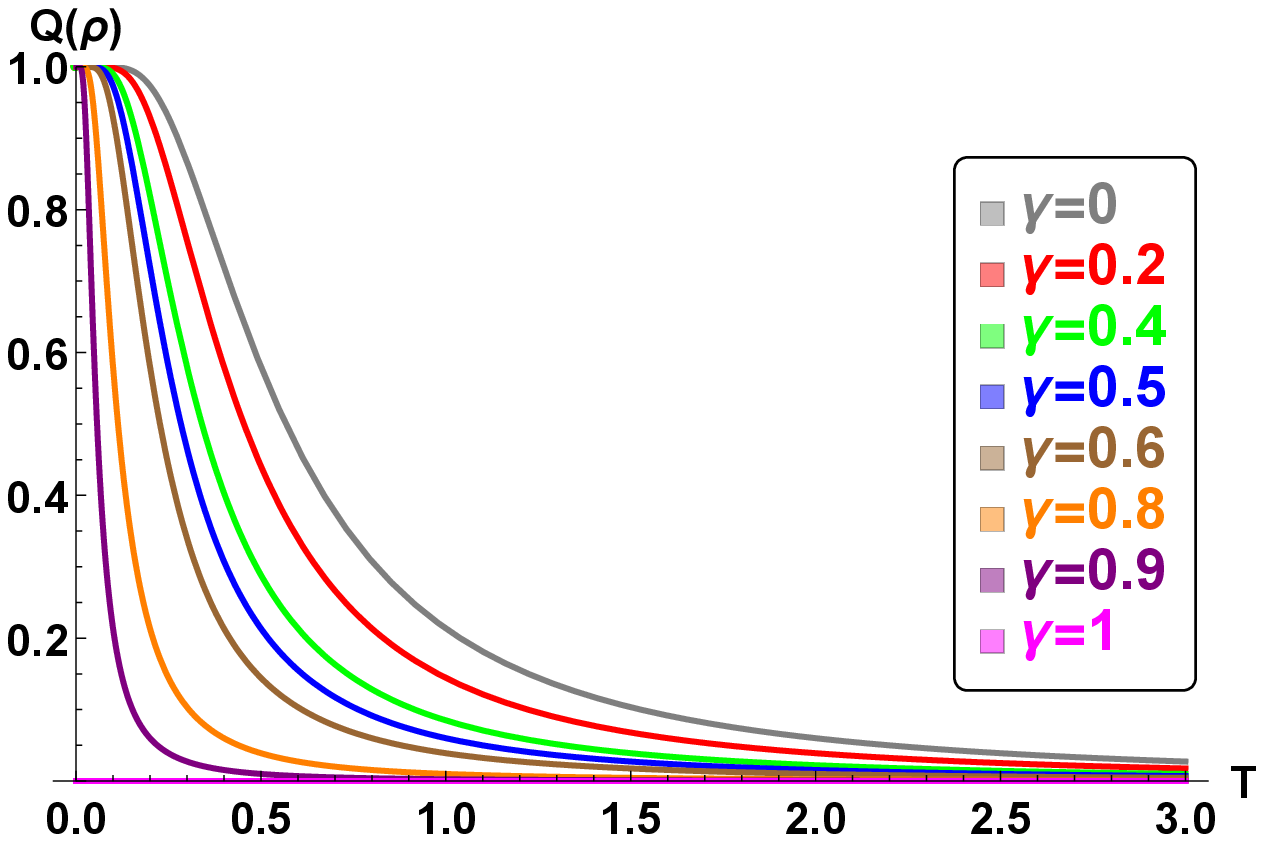}
\end{subfigure}
    \caption{\sf Local quantum Fisher information versus the coupling parameter $\gamma$ and the temperature $T$.}\label{fig_1}
\end{figure}
\begin{figure}[H]
    \centering
    \begin{subfigure}[]
        \centering
        \includegraphics[width=3in]{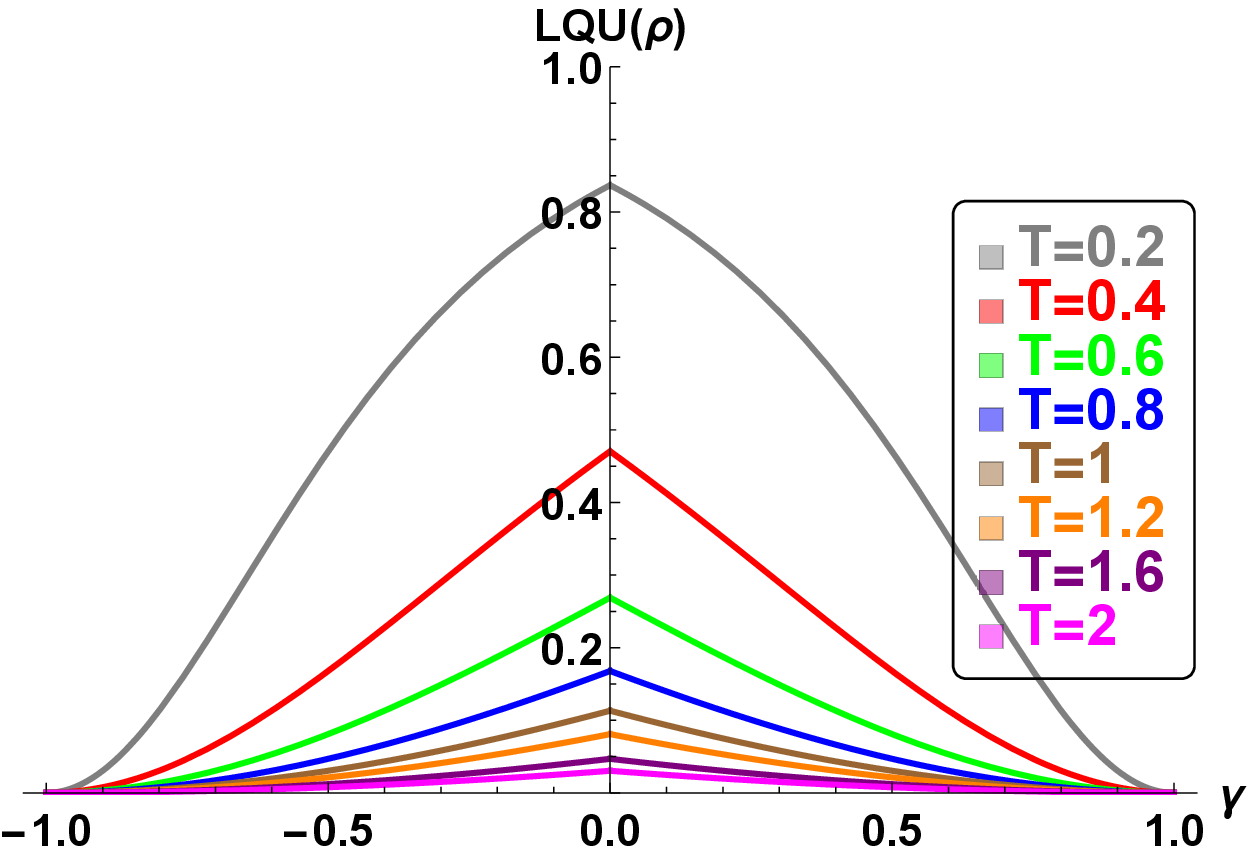}
    \end{subfigure}
    ~
    \begin{subfigure}[]
        \centering
        \includegraphics[width=3in]{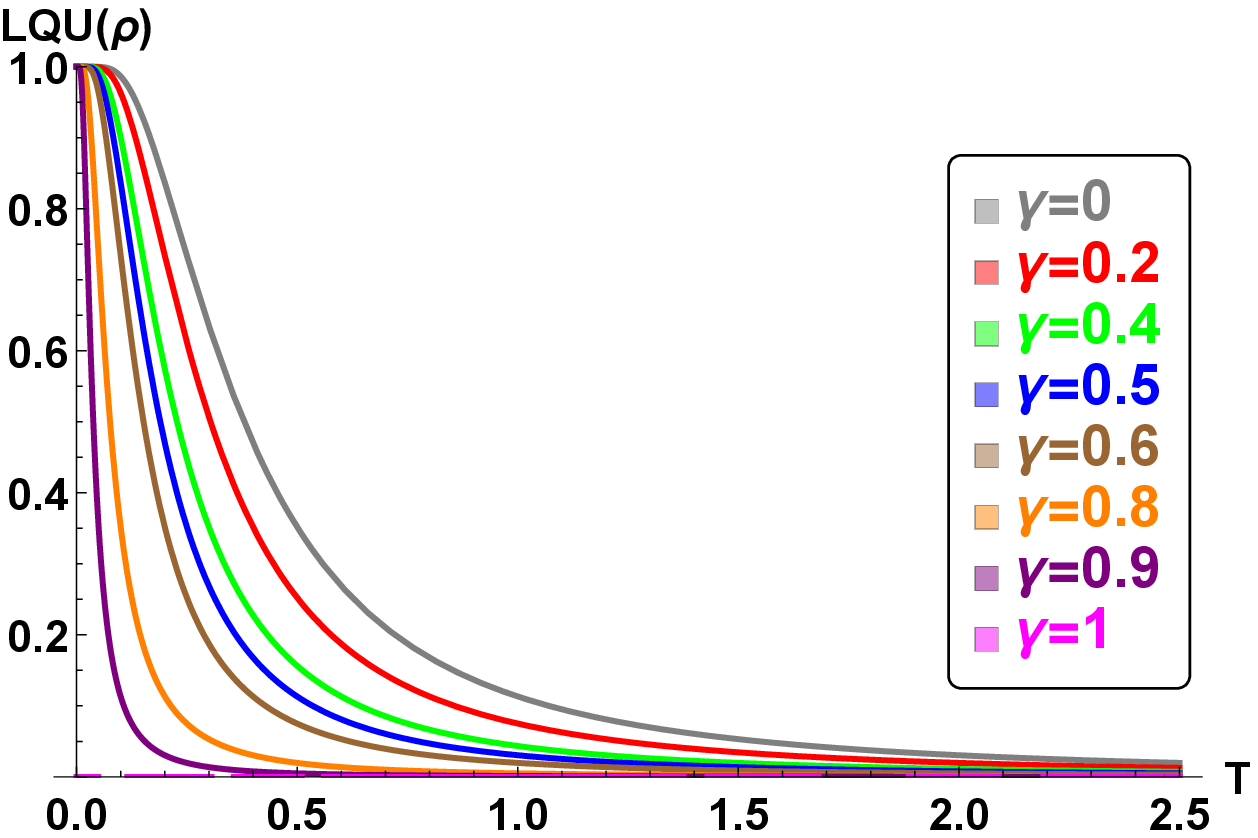}
    \end{subfigure}
    \caption{\sf Local quantum uncertainty versus the coupling parameter $\gamma$ and the temperature $T$.} \label{fig_2}
\end{figure}
In Fig.\ref{fig_1} (a), we depict the behavior of local quantum Fisher information versus the coupling parameter $\gamma$ for various values of the temperature. It is clearly seen from the results reported in this figure that the maximum amount of local quantum Fisher information is obtained when $\gamma\longrightarrow0$ and for low temperatures. This indicates that the isotropic $XX$ model exhibits more pairwise quantum correlations, especially for lower values of temperatures. We notice that increasing the temperature values tend to reduce the amount of quantum correlations in the system. In the limiting case when $\gamma\longrightarrow1$, which corresponds to Ising model, the quantum correlations vanish (even for low temperature). The results reported in Fig.\ref{fig_1}(b) confirm the results obtained in Fig.\ref{fig_1}(a). In particular, one verifies that the maximal amount of quantum correlations is exhibited by isotropic $XX$ model ($\gamma=0$) and the temperature effects tend to destroy the quantum correlations.\\
Let us now analyze the behavior of quantum correlations measured by local quantum uncertainty. As depicted in Fig.\ref{fig_2}(a) and Fig.\ref{fig_2}(b), it is obvious that the local quantum uncertainty and local quantum Fisher information exhibit similar variation versus $B$ and $T$. Furthermore, it is simple to see from the results depicted in Fig.1 and Fig.2 that the amount of quantum correlations measured by local quantum Fisher information is greater than the local quantum uncertainty. This result agrees with the inequality (\ref{UQ2U}) which states that LQFI is always greater than LQU.
\subsection{Two qubit isotropic $XY$ model with a magnetic field}
Consider now the Hamiltonian for the two qubit isotropic $XY$ model subjected to an external magnetic field $B$ along the $z$ axis. For $N = 2$, $J_x = J_y=J$ and $J_z = 0$, the Hamiltonian (\ref{H}) becomes
\begin{equation}
H = J\sum\limits_{i = 1}^2 {\left( {\sigma _i^x\sigma _{i + 1}^x + \sigma _i^y\sigma _{i + 1}^y} \right) + \frac{B}{2}\sum\limits_{i = 1}^2 {\sigma _i^z} }. \label{H2}
\end{equation}
In terms of the raising and lowering operators ${\sigma^\pm }$, the Hamiltonian (\ref{H2}) rewrites
\begin{equation}
H = J\left( {\sigma _1^ + \sigma _2^ -  + \sigma _1^ - \sigma _2^ + } \right) + \frac{B}{2}\left( {\sigma _1^z + \sigma _2^z} \right). \label{Hex2}
\end{equation}
The corresponding eigenvalues and eigenvectors are given by
\begin{equation}
H\left| {00} \right\rangle  = B\left| {00} \right\rangle, \hspace{0.5cm} H\left| {11} \right\rangle  =  - B\left| {11} \right\rangle, \hspace{0.5cm} H\left| {{\psi ^ \pm }} \right\rangle  = J\left| {{\psi ^ \pm }} \right\rangle,
\end{equation}
where $\left| {{\psi ^ \pm }} \right\rangle  = \frac{1}{{\sqrt 2 }}\left( {\left| {01} \right\rangle  \pm \left| {10} \right\rangle } \right)$ are the maximally entangled Bell states. In the standard computational basis, the density matrix of this system, in thermal equilibrium, is given by
\begin{equation}
\rho  = \frac{1}{Z}\left( {\begin{array}{*{20}{c}}
    {{e^{ - B\beta }}}&0&0&0\\
    0&{\cosh \left( {J\beta } \right)}&{ - \sinh \left( {J\beta } \right)}&0\\
    0&{ - \sinh \left( {J\beta } \right)}&{\cosh \left( {J\beta } \right)}&0\\
    0&0&0&{{e^{B\beta }}}
    \end{array}} \right),
\end{equation}
where the partition function in this case is given by $Z = \cosh \left( {J\beta } \right) + \cosh \left( {B\beta } \right)$. The eigenvalues and eigenvectors associated with this density matrix   are given respectively by
\begin{equation}
{p_1} = \frac{{{e^{ - B\beta }}}}{Z}, \hspace{1cm}{p_2} = \frac{{{e^{B\beta }}}}{Z},\hspace{1cm} {p_3} = \frac{{{e^{ - J\beta }}}}{Z},\hspace{1cm} {p_4} = \frac{{{e^{J\beta }}}}{Z},
\end{equation}
and
\begin{equation}
\begin{array}{*{20}{l}}
{\left| {{\psi _1}} \right\rangle  = \left| {00} \right\rangle ,\hspace{3cm}\left| {{\psi _2}} \right\rangle  = \left| {11} \right\rangle },\\
{\left| {{\psi _3}} \right\rangle  = \frac{1}{{\sqrt 2 }}\left( {\left| {01} \right\rangle  + \left| {10} \right\rangle } \right),\hspace{1cm}\left| {{\psi _4}} \right\rangle  = \frac{1}{{\sqrt 2 }}\left( {\left| {10} \right\rangle  - \left| {01} \right\rangle } \right).}
\end{array}
\end{equation}
Reporting these eigenvalues and eigenvectors in Eq. (\ref{Mlk}), it is simple to check that the matrix ${\rm M}$ is diagonal (i.e., ${{\rm M}_{12}} = {{\rm M}_{13}} = {{\rm M}_{23}} = 0$) and the diagonal elements are given by
\begin{equation}
{{\rm M}_{11}}={{\rm M}_{22}} = \frac{{2\left( 1 + \cosh\left( B\beta\right) \cosh \left( {J\beta }\right)\right)  }}{{{{\left( {\cosh \left( {B\beta } \right)+ \cosh\left( {J\beta }\right)} \right)}^2}}},
\end{equation}
and
\begin{equation}
    {{\rm M}_{33}} = \frac{{{\rm sech}\left({J\beta } \right)}}{{4\left( {\cosh\left({B\beta }\right)+ \cosh\left({J\beta } \right)} \right)}}.
\end{equation}
It is also simple to verify that
\begin{equation}
{{\rm M}_{11}} - {{\rm M}_{33}} = \frac{{7 + \cosh \left( {B\beta } \right)\left( {4\cosh \left( {2J\beta } \right) + 3} \right){\mathop{\rm sech}\nolimits} \left( {J\beta } \right)}}{{4{{\left( {\cosh \left( {B\beta } \right) + \cosh \left( {J\beta } \right)} \right)}^2}}}\geq 0.
\end{equation}
Using the expression of LQFI (\ref{LQFI}), one finds
\begin{equation}
\mathcal{Q}\left( \rho  \right) = 1 - \frac{{2\left( {1 + \cosh\left( {B\beta }\right)\cosh\left( {J\beta } \right)} \right)}}{{{{\left( {\cosh\left({B\beta }\right)+ \cosh\left({J\beta }\right)} \right)}^2}}}.
\end{equation}
On the other hand, to obtain the explicit expression of LQU, we compute first the elements of the matrix $W$ given by (\ref{w-elements}). The non-zero elements write as
\begin{equation}
{\omega_{11}} = {\omega_{22}} = \frac{{\sqrt {\left( {\cosh \left( {B\beta } \right) + 1} \right)\left( {\cosh \left( {J\beta } \right) + 1} \right)} }}{{\cosh \left( {B\beta } \right) + \cosh \left( {J\beta } \right)}},
\end{equation}
\begin{equation}
{\omega_{33}} = \frac{2}{{ {\cosh \left( {B\beta } \right) + \cosh \left( {J\beta } \right)} }}.
\end{equation}
Since ${\omega_{11}}\geq{\omega_{33}}$, LQU is simply given by
\begin{equation}
\mathcal{U}\left( \rho  \right) =1- \frac{{\sqrt {\left( {\cosh \left( {B\beta } \right) + 1} \right)\left( {\cosh \left( {J\beta } \right) + 1} \right)} }}{{\cosh \left( {B\beta } \right) + \cosh \left( {J\beta } \right)}},
\end{equation}
\begin{figure}[H]
    \centering
    \begin{subfigure}[]
        \centering
        \includegraphics[width=3in]{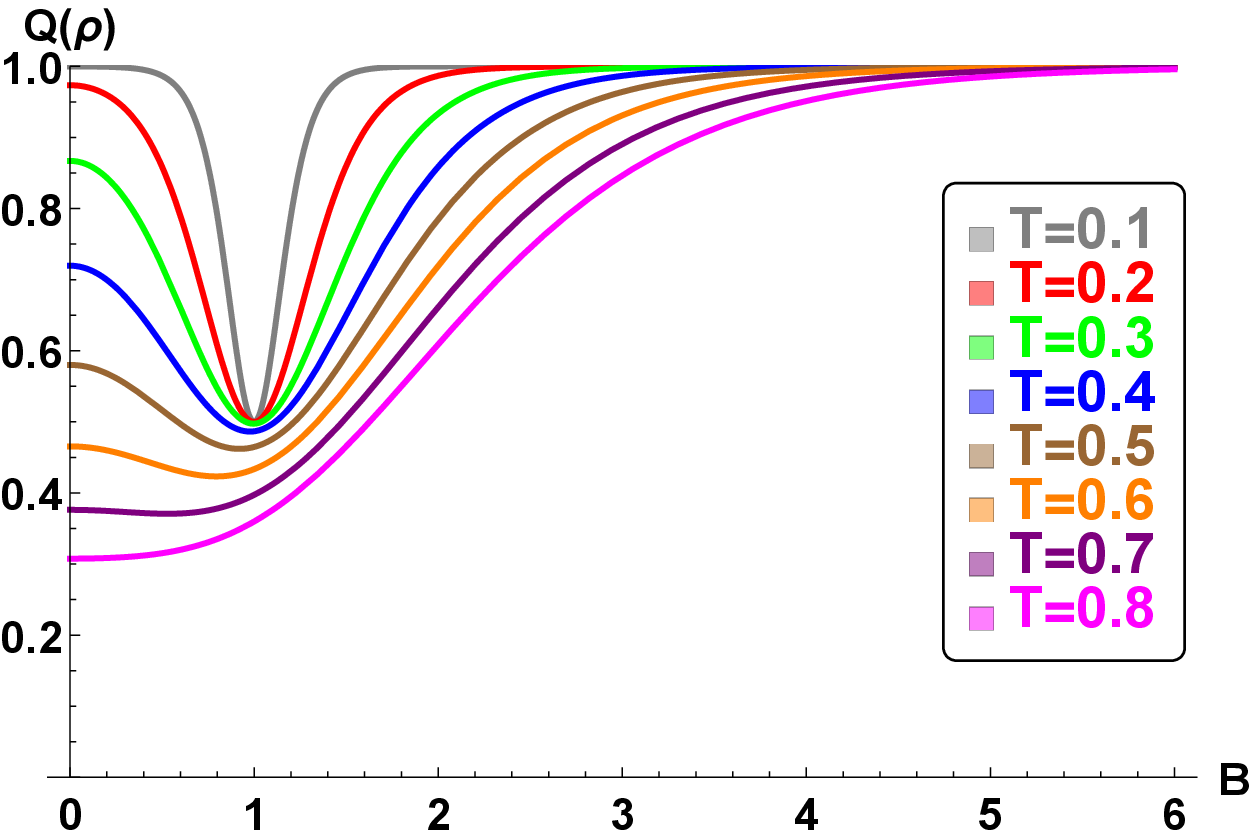}
    \end{subfigure}
    ~
    \begin{subfigure}[]
        \centering
        \includegraphics[width=3in]{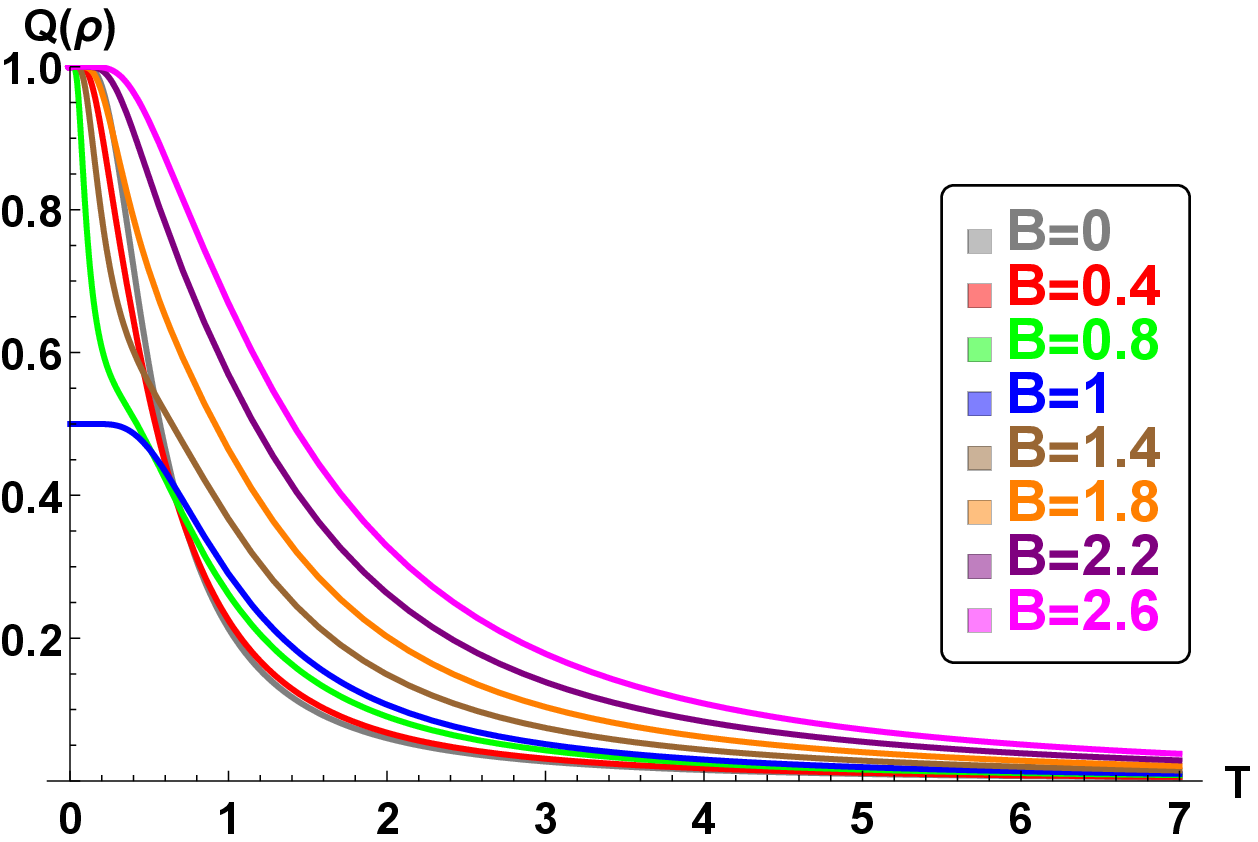}
    \end{subfigure}
    \caption{\sf Local quantum Fisher information versus the magnetic field $B$ and the temperature $T$.}\label{fig_3}
\end{figure}
\begin{figure}[H]
    \centering
    \begin{subfigure}[]
        \centering
        \includegraphics[width=3in]{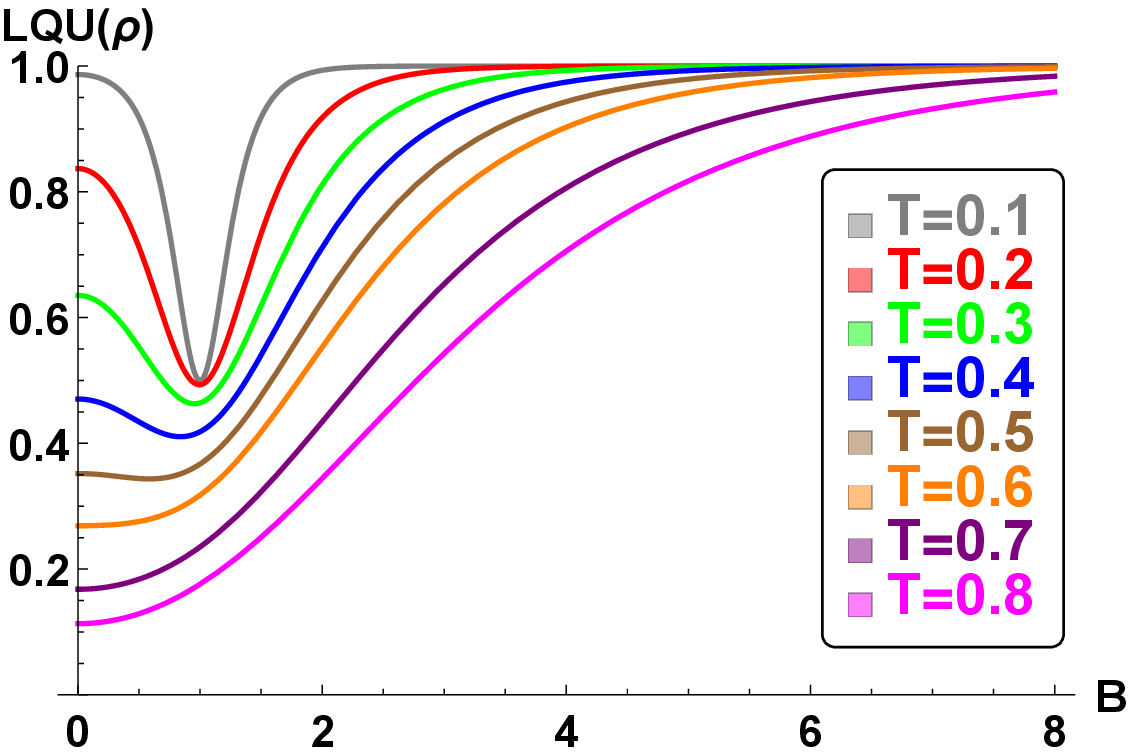}
    \end{subfigure}
    ~
    \begin{subfigure}[]
        \centering
        \includegraphics[width=3in]{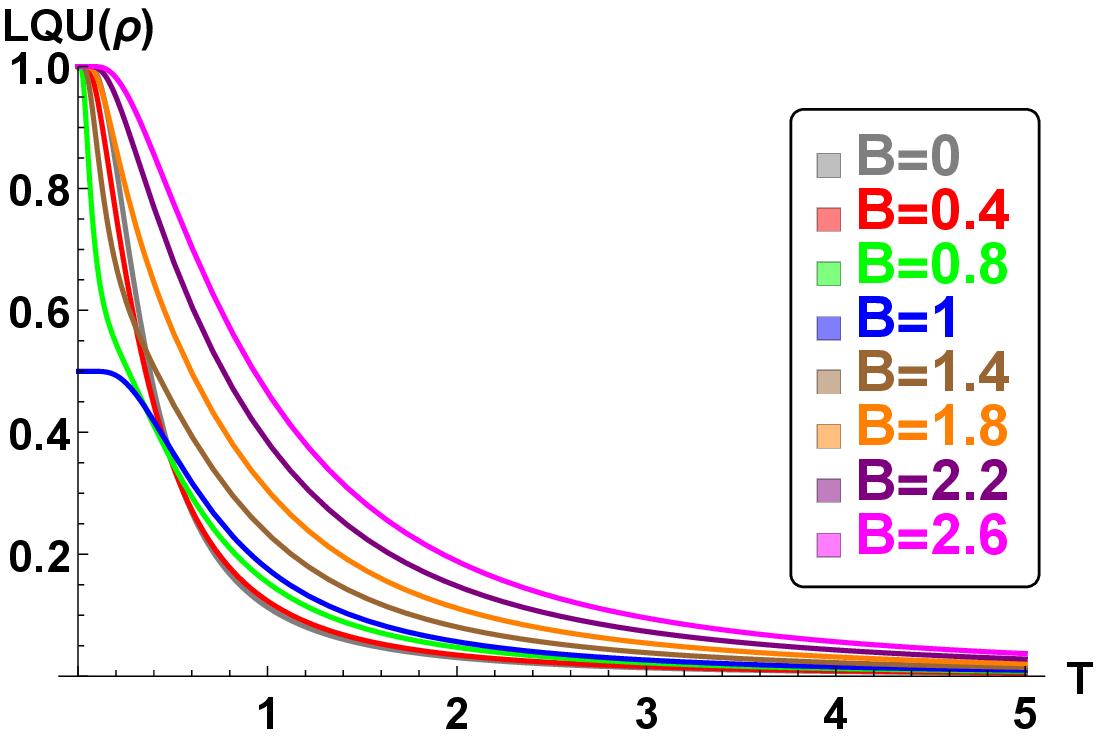}
    \end{subfigure}
    \caption{\sf Local quantum uncertainty versus the magnetic field $B$ and the temperature $T$.} \label{fig_4}
\end{figure}
The behavior of local quantum Fisher information in isotropic $XY$ model subjected to an external magnetic field is plotted in Figure \ref{fig_3}. Fig.\ref{fig_3}(a) gives the local quantum Fisher information versus the magnetic field $B$ for various values of the temperature $T$. The local quantum Fisher information exhibit a sudden change behavior for $B=1$. In addition, the increase of the magnetic field enhances the amount of quantum correlations contained in the system until it reaches a maximum amount of correlations. We notice that the amount of local quantum Fisher information is larger for low temperatures and it is reduced by increasing the values of $T$. The analysis of local quantum Fisher information versus the temperature reported in Fig.\ref{fig_3}(b) shows that the LQFI is maximal for $T=0$ and decreases monotonically with $T$ increasing to become zero for higher temperature. From Fig.\ref{fig_3}(a), the LQFI decreases for $B<1$ and low temperatures and starts increasing when $B>1$ to reach the maximal value $\mathcal{Q}(\rho)_{\rm max}=1$. This result is corroborated by the plot reported in Fig.\ref{fig_3}(b) which shows that for a fixed low temperature $T$, the quantum correlations decrease as $B$ increases. This shows that an external magnetic field destroys the amount of quantum correlation in the system for low temperature. On the other hand, the behavior of quantum correlations captured by local quantum uncertainty versus the magnetic field are depicted in Fig.\ref{fig_4}(a) and versus the temperature are reported in Fig.\ref{fig_4}(b). We remark that the local quantum Fisher information and the local quantum uncertainty exhibit similar variations versus the temperature $T$ and the magnetic field $B$ (see Figs \ref{fig_3} and \ref{fig_4}). The results reported in the figures \ref{fig_3} and \ref{fig_4} show that the local quantum Fisher information is always greater than the local quantum uncertainty. This agrees with the result given by the inequality (\ref{UQ2U}).

\section{Concluding Remarks}

In summary, the local quantum Fisher information plays an essential role in evaluating quantum correlations. This is due to its relationship with the concept of local quantum uncertainty. Also, the local quantum Fisher information is essential to determine the precision in metrological protocols.\\
To analyse the behaviors of these quantum correlation quantifiers, we have considered two variants of the Heisenberg $XY$ model. The first one concerns the anisotropic $XY$ model and the second variant deals with isotropic $XY$ model embedded in a magnetic field. Our results imply that the quantum correlations (LQFI or LQU) depends on the temperature and the coupling parameter in the anisotropic $XY$ model. The amount of quantum correlations take a maximum value for the isotropic $XX$ model ($\gamma=0$) and vanishes for the Ising model ($\gamma=\pm1$). In analysing the effect of temperature, we noticed that for higher values of temperature, the quantum correlations are destroyed. For a two qubit isotropic $XY$ model with a magnetic field, unlike the temperature, the increase of the magnetic field enhances the amount of quantum correlations contained in the system. Most importantly, the present investigation suggests that local quantum Fisher information and local quantum uncertainty exhibit a similar variation. This paper highlight the importance and interplay between these two special types of quantum correlations.

\end{document}